\documentclass[twocolumn]{cinc}
\pdfoutput=1
\usepackage{cite}
\usepackage{amsmath,amssymb,amsfonts}
\usepackage{algorithmic}
\usepackage{graphicx}
\usepackage{textcomp}
\begin{document}
	\bibliographystyle{cinc}
	
	% Keep the title short enough to fit on a single line if possible.
	% Don't end it with a full stop (period).  Don't use ALL CAPS.
	\title{A Mathematical Model for Skin Sympathetic Nerve Activity Simulation}
	
	% Both authors and affiliations go in the \author{ ... } block.
	% List initials and surnames of authors, no full stops (periods),
	%  titles, or degrees.
	% Don't use ALL CAPS, and don't use ``and'' before the name of the
	%  last author.
	% Leave an empty line between authors and affiliations.
	% List affiliations, city, [state or province,] country only
	%  (no street addresses or postcodes).
	% If there are multiple affiliations, use superscript numerals to associate
	%  each author with his or her affiliations, as in the example below.
	
	\author {Runwei Lin$^{1}$, Frank Halfwerk$^{2,3}$, Dirk Donker$^{4}$, Gozewijn Dirk Laverman$^{1,5}$, Ying Wang$^{1}$  \\
		\ \\ % leave an empty line between authors and affiliation
		$^1$ Biomedical signals and systems, University of Twente, the Netherlands \\
		$^2$ Engineering Organ Support Technologies, University of Twente, the Netherlands \\
		$^3$ Thorax Centrum Twente, Medisch Spectrum Twente, Enschede, The Netherlands \\
		$^4$ Cardiovascular and Respiratory Physiology, University of Twente, the Netherlands \\
		$^5$ Internal Medicine, Ziekenhuisgroep Twente, Almelo, the Netherlands}
	
	\maketitle
	
	% LaTeX inserts the ``Abstract'' heading in the proper style and
	% sets the text of the abstract in italics as required.
	\begin{abstract}
		Autonomic nervous system is important for cardiac function regulation. Modeling of autonomic cardiac regulation can contribute to health tracking and disease management. This study proposed a mathematical model that simulates autonomic cardiac regulation response to Valsalva Maneuver, which is a commonly used test that provokes the autonomic nervous system. Dataset containing skin sympathetic nervous activity extracted from healthy participants’ ECG was used to validate the model. In the data collection procedure, each participant was required to perform Valsalva Maneuver. The preliminary result of modeling for one subject is presented, and the model validation result showed that the root measure square error between the simulated and measured average skin sympathetic nervous activity is 0.01$\mu$V.  The model is expected to be further developed, evaluated using the dataset including 41 subjects, and ultimately applied for capturing the early signs of cardiac dysfunction in the future.  
	\end{abstract}
	% LaTeX inserts the extra space here automatically.
	
	\section{Introduction}
	Autonomic nervous system (ANS), consisting of sympathetic nervous system (SNS) and parasympathetic nervous system (PSNS), plays a major role in regulating cardiac function. Short term variation of cardiac autonomic tone change is mainly mediated by the baroreflex mechanism, which corresponds for the maintenance of blood pressure (BP) homeostasis. In particular, BP drop can lead to sympathetic arousal and increases the heart rate (HR). Despite the fact that HR and BP variation are related to the SNS, it is difficult to precisely quantify the change of cardiac sympathetic tone.
	
	Currently, the most widely adopted approaches for noninvasive SNS assessment are by monitoring changes of electrodermal activity (EDA) \cite{posada2020innovations} and heart rate variability (HRV) \cite{KANA2011251}. EDA measures the skin conductance variation caused by sweating, which has been proven to be solely controlled by SNS \cite{posada2020innovations}. Low frequency component of HRV is mediated by both SNS and PSNS and can therefore be used for evaluating cardiac sympathetic tone. In practice, EDA has higher sensitive than HRV in assessing SNS activity \cite{baghestani2024analysis}. Recently, Doytchinova et al. developed the skin sympathetic nervous activity (SKNA) measurement technology that can extract the SNS activity from traditional ECG measurement  \cite{DOYTCHINOVA201725}. A recent comparative study reports that SKNA has a quicker onset response even than EDA in assessing sympathetic nervous activity \cite{baghestani2024analysis}. Accordingly, monitoring the abnormal change of SKNA is expected to capture the very early signs of cardiac dysfunction. 
	
    However, it is challenging to quantify the changes of SKNA given its vulnerable signal quality, which hampers its application in daily monitoring. Physiology informed mathematical modeling has been proposed to simulate the behaviors of autonomic cardiovascular regulation \cite{KANA2011251,Ottesen2011}. We hypothesize that a SKNA mathematical model can provide better signal quantification and help disease management. Therefore, we proposed the first autonomic cardiac regulation model for SKNA simulation. The model will be used to detect the early signs of cardiac diseases in the future.  
	\section{Method}
	\subsection{Dataset description} 
	The experiment was approved by the Natural Science and Engineering Sciences ethical committee of University of UTwente with ethical number 2022.153. Forty-one healthy participants were recruited in  the experiment (21 female, 20 male). All participants signed the informed consent. The dataset was collected at the outpatient clinic of the Department of Cardiology, Medisch Spectrum Twente (MST), the Netherlands. Each participant took part in two 3-minute measurements. In both measurements, participants  were required to perform two 15-second VMs and rest for one minute in between. Signals containing SKNA and ECG were recorded using the Biomonitor (Mega Electronics Ltd., Kuopio, Finland) and the Einthoven's triangle configuration \cite{tertoolen2023evaluation}. 
	\subsection{Data preprocessing}
	SKNA has higher frequency and lower amplitude compared to ECG \cite{baghestani2024analysis}. Previous studies suggests that SKNA ranging from 500Hz to 1000Hz is most sensitive to the VM \cite{baghestani2024analysis}. SKNA signals were therefore extracted by a fourth order Butterworth filtering ranging [500Hz, 1000Hz]. Since SKNA has a relatively high frequency range, integration and a moving average method was performed before further analysis \cite{baghestani2024analysis}. We calculated the average of the integrated SKNA (aSKNA) within each 100-millisecond, which is in line with the previous study \cite{baghestani2024analysis}. The ECG components were separated using a fourth order Butterworth filter ranging [0.5Hz, 40Hz] for optimal R peak extraction. Pan-Thompkins method \cite{pan1985real} was used for the detection of R peaks, and beat-to-beat HR was calculated based on the duration of the R-R intervals. Derived HR data was resampled to 10Hz for the alignment with aSKNA. To visualize the trend of HR and aSKNA, we calculated the mean and standard deviation of the normalized data from all participants.
	
	\subsection{Model description}
\begin{figure}[htbp]
		\centering	\includegraphics[width=0.95\columnwidth]{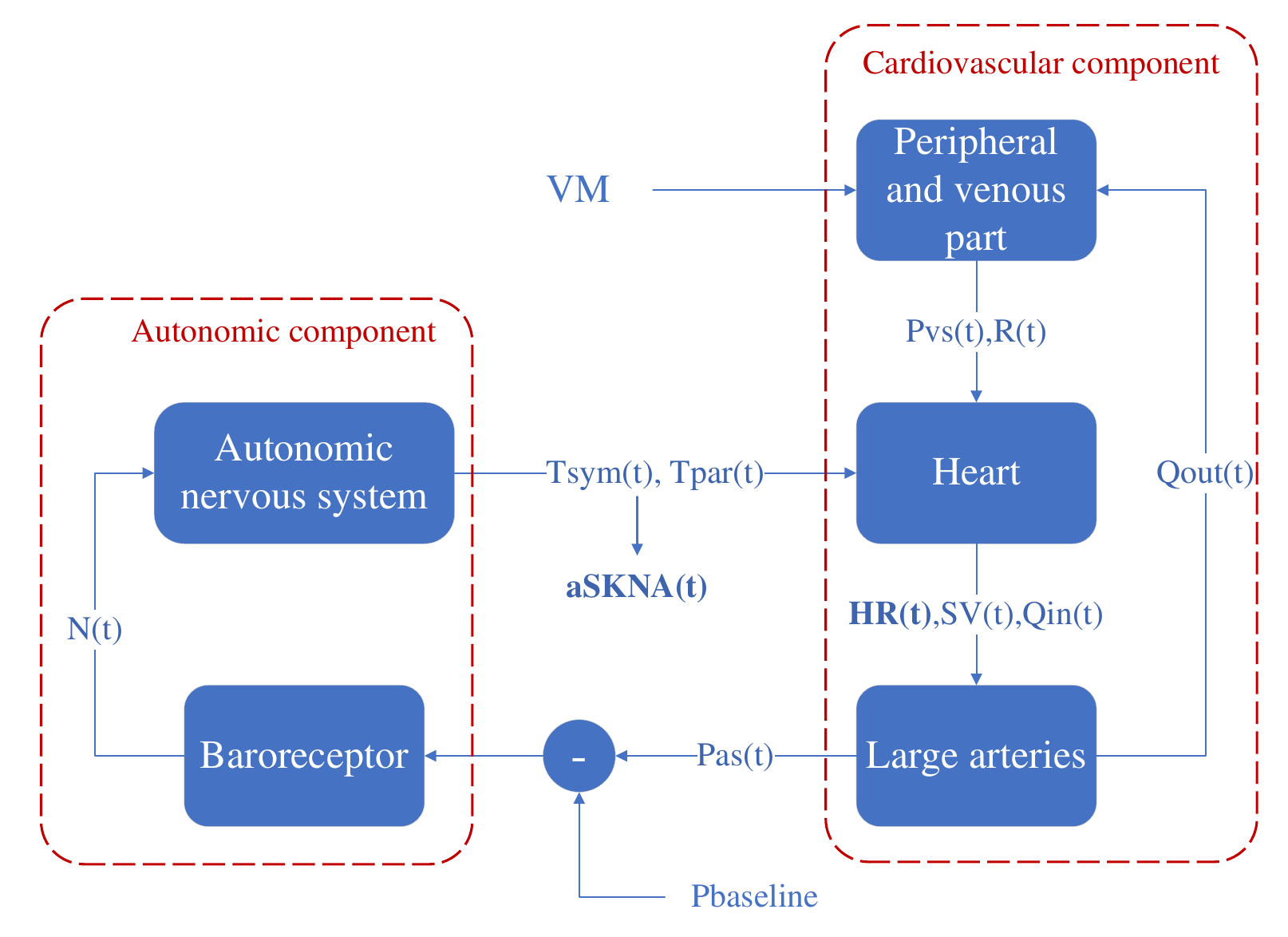}
		\caption{Scheme of the proposed autonomic cardiac regulation model.}
		\label{scheme}
	\end{figure}

	The model proposed in the study was developed based on the models proposed in \cite{KANA2011251, Ottesen2011}. Figure \ref{scheme} illustrates the model architecture.  
	The model was composed of an autonomic control component, cardiovascular system and a disturbance component. The autonomic component and cardiovascular system formed a closed-loop model that describes the change of mean arterial pressure [MAP; donated by $P_{as}(t)$ in the units of mmHg], and baroreceptor firing rate denoted by $N(t)$, in the unit of Hz. Change of MAP was characterized by the arterial compliance $C_{as}$ in the units of L/mmHg and difference between cardiac output ($Q_{in}(t)$ L/min) and the blood flow out of arteries ($Q_{out}(t)$ L/min) \cite{KANA2011251}, as shown in Eq. \ref{dpas}:
	\begin{equation}
		\frac{d P_{as}(t)}{dt}=\frac{1}{C_{as}}[Q_{in}(t)-Q_{out}(t)]
		\label{dpas}
	\end{equation}
	where $Q_{in}(t)$ was the multiplication between HR (bpm) and stroke volume (L/beat):
	\begin{equation}
		Q_{in}(t) = \text{HR}(t)*V_{str}(t)
	\end{equation}
	Stroke volume was modeled by the ratio of venous pressure $P_{vs}(t)$ (mmHg) over MAP \cite{KANA2011251}, as shown in Eq. \ref{vstr}:
	\begin{equation}
		\label{vstr}
		V_{str}(t) = k_{str}\cdot\frac{P_{vs}(t)}{P_{as}(t)}
	\end{equation}
	where $k_{str}$ is a coefficient in the unit of L/beat that describes the heart contractility. 
	$Q_{out}(t)$ was modeled from the difference between MAP and $P_{vs}(t)$ over the total peripheral resistance (TPR) R(t) (mmHg$\cdot$min/L) \cite{KANA2011251}:
	\begin{equation}
		Q_{out}(t) = \frac{P_{as}(t)-P_{vs}(t)}{R(t)}
	\end{equation}
	
	In the autonomic component, the baroreceptor firing rate N(t) altered with the difference between $P_{as}(t)$ and a baseline blood pressure ($P_{baseline}$ mmHg):
	\begin{equation}
		\frac{d N(t)}{d t}=(P_{as}(t)-P_{baseline}) \cdot N(t) \cdot k \cdot \frac{M-N(t)}{(M / 2)^2}-\frac{N(t)-N_0}{\tau}
		\label{dn}
	\end{equation}
	where $M$ is the maximum firing rate and $k$ (Hz/s$\cdot$mmHg) is a constant coefficient, $\tau$ (s) is the parameter that describes the response speed \cite{Ottesen2011}. Given the baroreceptor firing rate, Eq. \ref{tone} calculates the parasympathetic ($T_{par}(t)$) and sympathetic ($T_{sym}(t)$) tone as Ottesen proposed in \cite{Ottesen2011}:
	\begin{equation}
		\label{tone}
		T_{\text {par}}(t)=\frac{N(t)+N_0}{M}, \quad
		T_{\text{sym}}(t)=\frac{1-T_{par}(t)}{1+\beta \cdot T_{\text {par}}(t)}
	\end{equation}
	where $\beta$ is a weighting coefficient related to the inhibition of sympathetic tone by parasympathetic tone. 
	The aSKNA was only related with the sympathetic tone, we modelled it using a linear function:
	\begin{equation}
		\text{aSKNA}(t) = \text{aSKNA}_0 \cdot(1 + k_{askna} \cdot T_{sym}(t))
		\label{askna}
	\end{equation}
	In Eq. \ref{hr}, we described that HR(t) increases with the sympathetic tone ($T_{sym}(t)$) and inhibited by the parasympathetic tone ($T_{par} (t)$):
	\begin{equation}
		\text{HR}(t)=\text{HR}_0 \cdot\left(1+M_s T_{\text{sym}}(t)-M_p T_{\text{par}}(t)\right)
		\label{hr}
	\end{equation}
	where $M_s$ and $M_t$ are two positive constant coefficients and $\text{HR}_0$ is the intrinsic HR.
	
	Disturbance of the system is caused by VM. Based on the VM changing pattern in \cite{KANA2011251}, we divided and simplified the process of VM into three different phases: the expiration phase ($T_{begin}\leq t < T_{begin}+T_1$), the breath holding phase ($T_{begin}+T_1\leq t < T_{begin}+T_2$), and the release phase ($T_{begin}+T_2\leq t<\leq T_{begin} +T_3$), where $T_{begin}$, $T_1$, $T_2$, $T_3$ donated the VM onset time, duration of expiration phase, duration of breath holding phase and duration of release phase, respectively. Eq. \ref{pvs} showed two logistic functions we used to simulate the impact of VM on the venous pressure:
	\begin{equation}
		\label{pvs}    
		P_{vs}(t) =
		\begin{cases}   
			& P_{vs0} -k_{1}\frac{P_{s}/10\cdot e^{P_{rate1}\cdot (t-T_{begin})}}{P_{s}+(e^{P_{rate1}\cdot (t-T_{begin})}+1)}, \\
			&\quad\quad\quad \text{if } T_{begin} \leq t < T_{begin}+T_{1} 
			\\
			& P_{vs0} -k_{1}\frac{P_{s}/10\cdot e^{P_{rate1}\cdot T_1}}{P_{s}+(e^{P_{rate1}\cdot T_1}+1)}, \\
			&\quad\quad\quad \text{if } T_{begin} + T_1 \leq t < T_{begin}+T_{2}
			\\      
			& P_{vs0} + k_{2}\frac{P_{s}/10\cdot e^{P_{rate2}\cdot (t-T_{begin}-T_2)}}{P_{s}+(e^{P_{rate2}\cdot (t-T_{begin}-T_2)}+1)}, \\ 
			& \quad\quad\quad \text{if } T_{begin}+T_{2} \leq t \leq T_{begin}+T_{3} \\
			& P_{vs0}, \quad \text{if } t < T_{begin} \text{ or } t > T_3 \\
		\end{cases}
	\end{equation}
	where $P_{vs0}$ was the baseline venous pressure. $P_s$ is a constant of 40mmHg and $P_{rate1}$ and $P_{rate2}$ are two constant factors describing the changing rate of the logistic functions. $k_1$ and $k_2$ in the unit of mmHg are both scaling factors related to the impact of VM on venous pressure.
	\subsection{Model parameter identification}
	The differential equations (Eq. \ref{dpas} and Eq. \ref{dn}) were solved by modified Rosenbrock formula method of order 2 (Matlab function ode23s.m). To identify the model parameters, we chose the following cost function:
	\begin{equation}
		\begin{aligned}
			J(\gamma) = &\frac{1}{N}\sum\limits_{n=1}^{N}(|\frac{\text{HR}_{predict}(n)-\text{HR}_{measure}(n)}{{\text{HR}_{measure,max}}}|^2+ \\		&|\frac{\text{aSKNA}_{predict}(n)-\text{aSKNA}_{measure}(n)}{\text{aSKNA}_{measure,max}}|^2)
		\end{aligned}
	\end{equation}
	where $\gamma$ represents all the parameters to be identified. The cost function  J($\gamma$) calculated the average difference on all sample points between the measured and predicted metrics normalized by their max value from the measurement. It is minimized by the Neld-Melder algorithm, which is a commonly used non-gradient approach for biological system parameter identification. 
	\subsection{Model validation}
	The simulation of VM was generated based on the identified model parameters. The first VM of subject one was used for model validation. Root mean square error (RMSE) between the simulated result and measurement during the VM period was used as the metric for evaluation.
	\section{Results}
	\subsection{Dataset}
	Figure 2 shows the calculated mean and standard deviation of HR and aSKNA from all participants. Two VM period were highlighted from 60 second to 75 second, and from 135 second to 150 second, respectively. Both HR and aSKNA began to increase about 10 second before the VM began. During VM period, HR dropped firstly and then increased, yet aSKNA remained at an approximate constant level. They both returned back to the baseline about 5 seconds after the VM ended.  
	\begin{figure}[htb]
		\centering \includegraphics[width=0.87\columnwidth]{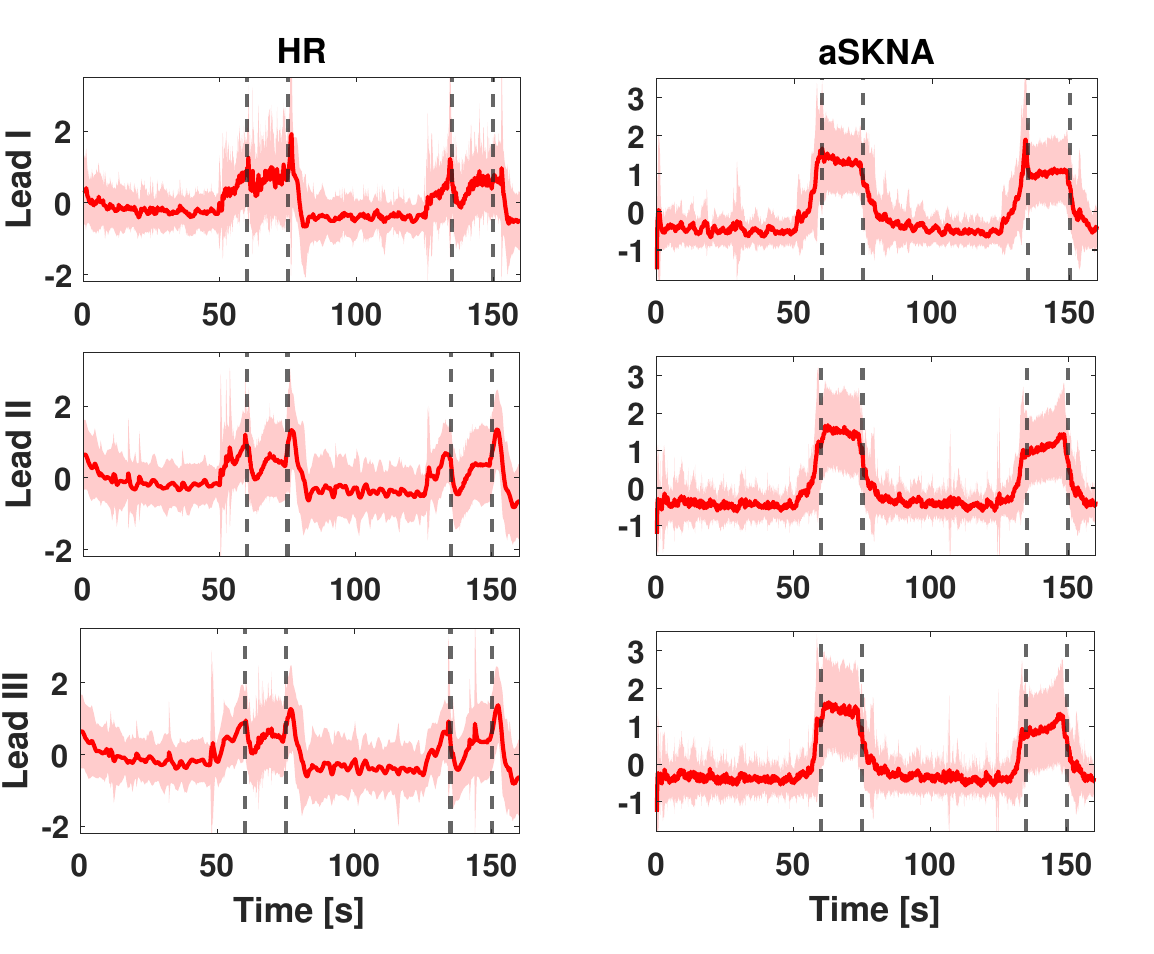}
		\caption{Mean and standard deviation of HR and aSKNA calculated from the three leads of all subjects. Dash lines indicates two VMs.}
		\label{fig:dataset}
	\end{figure}
	\subsection{Model validation}
	Figure 3 shows the validation result of a VM for one subject. The VM started approximately from 60-second and ended at 75-second. After VM began, both simulated HR and aSKNA increased to a constant level and dropped to baseline after the VM ended. The RMSE of HR is 5.45 bpm and the RMSE of aSKNA is 0.01$\mu$V.
	\begin{figure}[htbp]
		\centering
		\includegraphics[width=0.8\columnwidth]{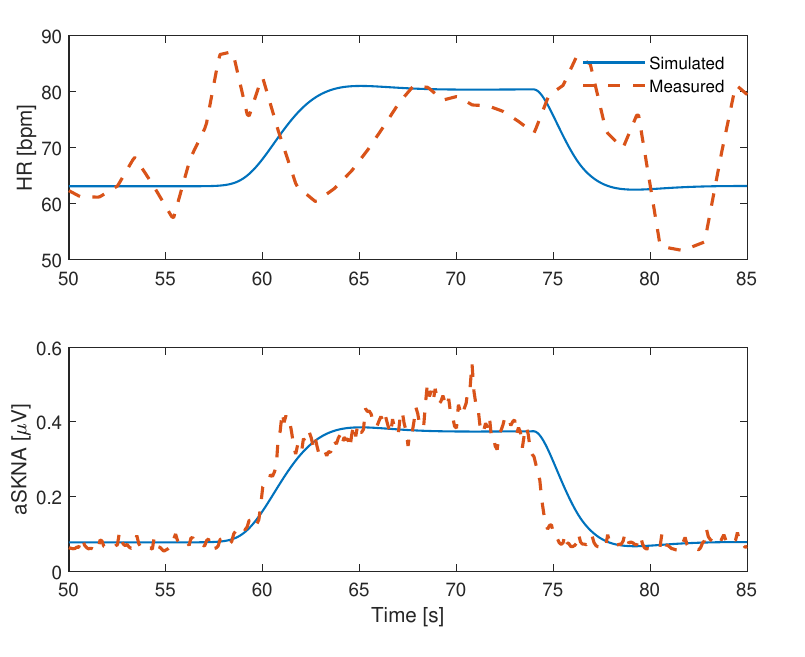}    
		\caption{An example of comparison between simulated and measured HR and aSKNA for one subject in the first VM. Blue curves are simulations and red dot curves are measurements.}
		\label{resulteg}
	\end{figure}
	\section{Discussion}
	This study proposed a mathematical model to simulate healthy participant’s autonomic cardiac regulation. The model is closed-loop and consists of autonomic and cardiovascular compartments. Model outputs are HR and aSKNA, two variables that can be used to estimate the cardiac sympathetic tone. Performance of the model was validated on ECG and SKNA recorded from healthy participants. The preliminary result of one subject showed that the model provides proper estimation of HR and aSKNA response to VM.
	
	Figure \ref{fig:dataset} shows that both average aSKNA and HR increase above baseline during the VM period, proving that aSKNA can be a marker for evaluating SNS activity. The simulated HR and aSKNA were close to the measurement in the VM period. The preliminary result showed that the RMSE between simulation and measurement is 0.01$\mu$V for the aSKNA and 5.45bpm for HR, which proved the appropriate model design and parameter selection. However, the model also has certain limitations. The first limitation is the model neglects the impact of VM induced peripheral resistance change \cite{KANA2011251}. In addition, current model does not incorporate sympathetic delay \cite{Ottesen2011}. Parameters such as arterial compliance was fixed for stability, however, it might change over time in reality. We also simplified the changing pattern of VM, which might account for the deviation between prediction and the measurement. Future work will improve the parameter identification accuracy using machine learning techniques.
	
	Despite its frequent usage in autonomic cardiac regulation assessment, it is not feasible to apply VM test in daily life scenarios. Physiological responses during daily activities, such as sleep or exercise, are closely related to the function of ANS. Therefore, we expect that the model presented in this paper can be further developed and applied to the daily monitoring of the ANS function. In addition, the model is also expected to embrace wearable digital markers such as photoplethysmogram or EDA signals to provide valuable insights in relevant diseases detection and prevention, such as cardiovascular diseases, diabetes and sleep disorders.
	
	\section{Conclusion}
	This work proposed the first physiology informed mathematical model to simulate the average SKNA. The model was validated on SKNA and ECG data collected while healthy participants performing VM. Results showed that the model could properly estimate HR and aSKNA in VM. Future work will extend the model to capture the early signs of cardiovascular dysfunction in daily monitoring applications.
	
	\section*{Acknowledgments}  
	% This section is not numbered.
	% 
The authors would like to acknowledge Dr. Vincent van der Pas, Jacomine Tertoolen among other students for data collection. 
	
	\bibliography{refs}
	
	% If you don't use BibTeX (why not?) , comment out or remove the previous
	% line, and uncomment the following lines up to the ``}\end{bibliography}''
	% line below:
	%\begin{thebibliography}{99}{ %\small
	% \bibitem{tag} (General form) J. K. Author, ``Name of paper,''
	%   \emph{Abbrev. Title of
%   Periodical}, vol. x, no. x, pp. xxx--xxx, Abbrev. Month, year. 

% \bibitem{ito}  M. Ito et al., ``Application of amorphous oxide TFT to
%   electrophoretic display,'' \emph{J. Non-Cryst. Solids}, vol. 354, no. 19,
%   pp. 2777--2782, Feb. 2008.

% \bibitem{fardel}  R. Fardel, M. Nagel, F. Nuesch, T. Lippert, and
%   A. Wokaun, ``Fabrication of organic light emitting diode pixels by
%   laser-assisted forward transfer,'' \emph{Appl. Phys. Lett.}, vol. 91,
%   no. 6, Aug. 2007, Art. no. 061103.

% \bibitem{buncombe} J. U. Buncombe, ``Infrared navigation Part I: Theory,''
%     \emph{IEEE Trans. Aerosp. Electron. Syst.}, vol. AES-4, no. 3,
%     pp. 352--377, Sep. 1944.

% Uncomment the following line if you are not using BibTeX.
%}\end{thebibliography}

% LaTeX inserts the ``Address for correspondence'' heading.
\begin{correspondence}
Runwei Lin, Ying Wang\\
Drienerlolaan 5, 7522 NB Enschede, The Netherlands\\
r.lin@utwente.nl, imwywk@gmail.com
\end{correspondence}

\end{document}